\documentclass[preprint,12pt,sort&compress,nopreprintline]{elsarticle}

\makeatletter
  \long\def\pprintMaketitle{\clearpage
  \iflongmktitle\if@twocolumn\let\columnwidth=\textwidth\fi\fi
  \resetTitleCounters
  \def\baselinestretch{1}%
  \printFirstPageNotes
  \begin{center}%
 \thispagestyle{pprintTitle}%
   \def\baselinestretch{1}%
    \Large\@title\par\vskip18pt
    \normalsize\elsauthors\par\vskip10pt
    \footnotesize\itshape\elsaddress\par\vskip10pt
    \end{center}%
  \gdef\thefootnote{\arabic{footnote}}%
  }
\makeatother
\usepackage{siunitx}
\usepackage{amsmath}
\usepackage{graphicx}
\usepackage{float}
\usepackage{subfigure}
\usepackage{epstopdf}
\usepackage{color}
\usepackage{multirow}
\usepackage{setspace}
\usepackage{overpic}
\usepackage{amssymb}
\usepackage[colorlinks,urlcolor=blue, citecolor=blue,linkcolor=blue] {hyperref}
\usepackage{lineno}
\usepackage{bm}
\usepackage{rotating}
\usepackage{array}
\usepackage{booktabs}
\usepackage{besphysics}
\usepackage[utf8]{inputenc}
\usepackage{enumitem}
\renewcommand{\thefootnote}{\fnsymbol{footnote}}

\newcommand{\decaychainc}{D_s^{*+} \to D_s^{+}(\to K^{+} K^{-} \pi^{+})\pi^{0}}
\newcommand{\BESIIIorcid}[1]{\href{https://orcid.org/#1}{\hspace*{0.1em}\raisebox{-0.45ex}{\includegraphics[width=1em]{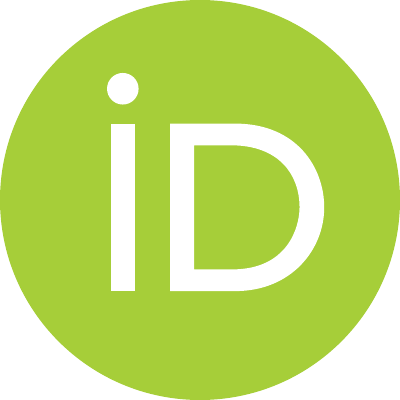}}}}

\let\oldequation\equation
\let\oldendequation\endequation
\renewenvironment{equation}
 {\linenomathNonumbers\oldequation}
 {\oldendequation\endlinenomath}
 
\begin{document}
\journal{Science Bulletin}

\begin{frontmatter}
\title{\boldmath Precision Measurement of the $D_{s}^{*+} - D_{s}^{+}$ Meson Mass Difference}
	\author{BESIII Collaboration\footnote{Corresponding author (email: besiii-publications@ihep.ac.cn)}\footnote{Authors are listed at the end of this paper.}}
\end{frontmatter}

\noindent
Received: 11-Mar-2026\\
Revised: 13-Apr-2026\\
Accepted: 21-Apr-2026\\

High precision measurements of charmed meson masses provide stringent tests of 
theoretical frameworks such as the chiral perturbation theory ($\chi$PT) that underpin our understanding of quantum chromodynamics (QCD) 
in the non-perturbative regime (See Ref.~\cite{Meng:2022ozq} for a recent review). 
While the mass differences $\Delta m_{+} \equiv m_{D^{*+}} - m_{D^+}$ and 
$\Delta m_{0} \equiv m_{D^{*0}} - m_{D^0}$ have uncertainties of
15~keV/$c^2$ and 30~keV/$c^2$ respectively~\cite{pdg}, the corresponding 
charm-strange quantity 
$\Delta m_s \equiv m_{D_s^{*+}} - m_{D_s^+} = (143.8 \pm 0.4)$~MeV/$c^2$~\cite{pdg} 
has an uncertainty of a full order of magnitude greater. 
This precision gap presents a significant obstacle for testing 
$\chi$PT predictions that include light quark mass effects and 
electromagnetic corrections~\cite{Goity:2007fu,Karliner:2019lau,Meng:2022ozq}, for validating lattice QCD calculations 
of heavy-light meson properties~\cite{Donald:2013sra}, for understanding the nature 
of the $D^{*}_{s0}(2317)-D_{s1} (2460)$ doublet~\cite{Alhakami:2019ait}, and recently discovered exotic states whose masses lie within a few MeV/$c^2$ of 
$D_s^{*+}\bar{D}$ thresholds~\cite{BESIII:2020qkh,BESIII:2022qzr}.

To measure $\Delta m_s$, the isospin-breaking decay $D_s^{*+}\to D_s^+\piz$ is chosen owing to the very low Q-value in the transition, as compared to the dominant radiative decay $D_s^{*+}\to D_s^+\gamma$. The lower Q-value leads to a considerably narrower line shape of the difference between the reconstructed $D_s^{*+}$ and $D_s^{+}$ masses, $\Delta M_s\equiv M(D_s^{+}\pi^0)-M(D_s^+)$, as shown in the Supplementary material.

A critical experimental challenge in measuring $\Delta m_s$ then arises from the 
systematic uncertainties associated with reconstructing the very soft $\pi^0$ mesons 
produced in $D_s^{*+} \to D_s^+ \pi^0$ decays, which have laboratory momenta of 
only $\sim$60~MeV/$c$. Electromagnetic calorimeter responses at such low energies 
are difficult to model accurately in Monte Carlo (MC) simulations, leading to potential biases 
in $\pi^0$ momentum reconstruction that directly propagate into  mass difference 
measurements. Previous measurements~\cite{CLEO:1995ewe} relied on detector simulations 
rather than data-driven calibrations, limiting achievable precision. 

In this work, we introduce a novel $\pi^0$ momentum calibration technique that 
overcomes this limitation, and report a precise measurement of the $D_s^{*+}-D_s^+$ mass difference using 3.19~$\ifb$ $\ee$ annihilation data collected at a center-of-mass (c.m.) energy of $E_{\rm c.m.}=4.178\gev$ with the BESIII detector in 2016. 
Using the kinematically similar decay $D^{*+} \to D^+ \pi^0$ 
as a control channel, we calibrate the $\pi^0$ energy response in two-dimensional 
bins of momentum magnitude $p(\pi^0)$ and polar angle $\cos\theta(\pi^0)$, anchoring 
the calibration to the precisely known $D^{*+} - D^+$ mass difference $\Delta{m_+}$~\cite{pdg}. 
With this method, an SU(3) flavor violating parameter, $\Delta m_D\equiv\Delta m_s-\Delta{m_+}$~\cite{Goity:2007fu,Karliner:2019lau,Meng:2022ozq}, 
is also obtained with the external uncertainty of $\Delta{m_+}$~\cite{pdg} canceled in the difference. 
This data-driven approach can be applied to any measurement involving soft 
$\pi^0$ reconstruction, {\it e.g.}, measuring masses of $D^{*}_{s0}(2317)$ and $D_{s1}(2460)$
 through the final states of $D_s^+\pi^0$ and $D_s^{*+}\pi^0$, respectively. 
 
Details about the design and performance of the BESIII and MC simulation are given in Refs.~\cite{Ablikim:2009aa, geant4, ref:kkmc1, ref:kkmc2, ref:evtgen1, ref:evtgen2, ref:lundcharm1, ref:lundcharm2}. 
The inclusive MC sample includes the production of $D_s^{*+}$ 
particles via
the process $e^+e^-\to D_s^{*\pm}D_s^{\mp}$,
and other open charm
processes, as well as the initial state radiation production of vector charmonium(-like) states,
and the continuum processes incorporated in {\sc
kkmc}~\cite{ref:kkmc1, ref:kkmc2}.
Subsets of the inclusive MC sample that include \mbox{$D^{*+}_{(s)}\to D_{(s)}^+\pi^0$} decays are generated with a vector-to-two-scalars model~\cite{ref:evtgen1, ref:evtgen2}, 
and used as dedicated MC samples to simulate our signal decays. 

We study the $D_s^{*+}\rightarrow D_s^+ \pi^0$ decay with the cascade decay $D_s^+\to K^+ K^- \pi^+$ to determine the difference between the $D_s^{*+}$ and $D_s^{+}$ masses $\Delta m_s$. Charge-conjugation modes are implied throughout this work. The distribution of $\Delta M_s$, is fitted to extract $\Delta m_s$. The signal component in the $\Delta M_{s}$ fit is a resolution function determined from our MC simulation of the detector response. A key factor of having an unbiased measurement on $\Delta m_s$ is to model the reconstruction of $\pi^0$ momentum precisely in the simulation. 
This is achieved by correcting reconstructed $\pi^0$ kinematics in MC simulation using dedicated calibration sample of \mbox{$D^{*+}\to D^+ (\to K^- \pi^+ \pi^+) \pi^0$} decays, so that the measured central value of the $D^{*+}-D^{+}$ mass difference $\Delta m_+$ coincides with the current world average~\cite{pdg}.  The correction factors are then validated using a sample of $D^{*+}\to D^+ (\to K^+ K^- \pi^+) \pi^0$ decays.

The event selection criteria for the decays of $D_{(s)}^{*+}\to D_{(s)}^+\pi^0$ are based on the reconstruction of one three-prong $D_{(s)}^+$ candidate and 
one $\pi^0$.
This analysis  uses the same selection criteria of $K^\pm$, $\pi^\pm$, and $\pi^0$ mesons as in Refs.~\cite{BESIII:2020qkh,BESIII:2021jnf,BESIII:2024thr}. In the $\pi^0 \to \gamma\gamma$ reconstruction, the diphoton invariant mass $M(\gamma\gamma)$ is additionally required to be between $120$~and~$150$~$\mevcc$.
A kinematic fit that constrains $M(\gamma\gamma)$ to the known $\pi^0$ mass~\cite{pdg} is performed on the selected photon pairs. This  significantly improves the reconstructed $\pi^0$ momentum resolution to $\sigma_p/p\sim2$\%. Due to the small Q-value in the $D_{(s)}^{*+}\to D_{(s)}^+\pi^0$ decays, the $\pi^0$ momentum is required to be less than 100~MeV/$c$ to suppress more energetic $\pi^0$s from decays of ground-state charmed mesons and light hadrons. In case of multiple $\pi^0$ candidates in an event, only the one with $M(\gamma\gamma)$ closest to the known $\pi^0$ mass is selected.

For each $D_{(s)}^+$ candidate, 
the three-track invariant mass $M(D_{(s)}^{+})$ is required to satisfy \mbox{$|M(D_{(s)}^{+})-m_{D_{(s)}^+}|<12~\mevcc$}, where $m_{D_{(s)}^+}$ is the known mass of $D_{(s)}^+$~\cite{pdg}.
The reconstructed $D_{(s)}^{+}$ mass distributions in data for the calibration and signal channels are shown in the Supplementary material.
The following requirements are imposed 
on $D_{(s)}^+$ and $D_{(s)}^{*+}$ candidates to ensure the $D_{(s)}^+$ originates from $D_{(s)}^{*+}$ and improve signal purity.
We define the recoil mass of $D^{(*)+}_{(s)}$ as

 \begin{align}
M_{\rm rec}&(D_{(s)}^{(*)+})c^2 \equiv \nonumber\\
 &\sqrt{\left(E_{\rm c.m.} - \sqrt{\left|\vec{p}_{D_{(s)}^{(*)+}}\right|^2c^2+m^2_{D_{(s)}^{(*)+}}c^4}\right)^2-\left|\vec{p}_{D_{(s)}^{(*)+}}\right|^2c^2}\,,
 \end{align}

\noindent where $\vec{p}_{D_{(s)}^{(*)+}}$ is the reconstructed three-momentum of the $D_{(s)}^{(*)+}$ candidate in the $e^+e^-$ c.m. frame, and $m_{D_{(s)}^{*+}}$ is the known mass of $D_{(s)}^{*+}$~\cite{pdg}. 
We require \mbox{$\left|M_{\rm rec}(D^{+})-m_{D^{*+}}\right|> 0.1$~\gevcc\ }to suppress backgrounds of $D^+$ mesons from the process \mbox{$e^+e^-\to D^{+}D^{*-}$}. 
Furthermore, we require the mass difference \mbox{$\left|M_{\rm rec}(D^{+})-M_{\rm rec}(D^{*+})-\Delta m_+^{\rm PDG}\right|<3$~\mevcc\ }to suppress background events with $\pi^0$ mesons from the process \mbox{$e^+e^-\to D^{+}D^{*-}(\to D^-\pi^0)$}.
As $D^{*+}$ particles are predominantly produced in the processes of $e^+e^-\to D^{*+}D^{-}$ and $e^+e^-\to D^{*+}D^{*-}$, we require 
$M_{\rm rec}(D^{*+})\in ([m_{D^+}-20,\ m_{D^+}+20]\cup[m_{D^{*+}}-20,\ m_{D^{*+}}+20])$~\mevcc, to be around the known mass of $D^-$ or $D^{*-}$. 
Similarly, as $D_s^{*+}$ particles are solely produced in the process $e^+e^-\to D_s^{*+}D_s^{-}$, we require $M_{\rm rec}(D_s^{*+})\in [m_{D_s^+}-15,\ m_{D_s^+}+15]$~\mevcc\ to suppress background from random combinations of $\pi^0$ and $D_s^+$ mesons. 

As in Ref.~\cite{DSTpDpbabar2017}, the distribution of the raw difference between the reconstructed $D_s^{*+}$ ($D^{*+}$) and $D_s^{+}$ ($D^{+}$) masses, $\Delta M_s$ ($\Delta M_+$), is fitted to extract $\Delta m_s$ ($\Delta m_+$). While the background shapes are modeled with the signal-removed inclusive MC sample, using the kernel-estimation method~\cite{Cranmer:2000du}, the signal component in the $\Delta M_i,\, (i\in \{+,s\})$ 
fit is a 
sum of three 
Probability Density Functions (PDFs),

\begin{align}
{\cal S}(\Delta  &M_{i}) = f_1 \cdot {\rm G}(\Delta M_{i}; \Delta m_{i}+\delta_{\Delta m_{i}}, \varepsilon\cdot\sigma_1) \nonumber \\
& + (1-f_1)\cdot\left[f_2\cdot {\rm CB}(\Delta M_{i};\Delta m_{i}+\delta_{\Delta m_{i}}, \varepsilon\cdot\sigma_2, \alpha, n)\right.  \nonumber \\
& \left. + (1-f_2) \cdot{\rm BfG}(\Delta M_{i};\Delta m_{i}+\delta_{\Delta m_{i}}, \varepsilon\cdot\sigma^{L}_3,\varepsilon\cdot\sigma^{R}_3)\right]\,,
        \label{eq:sigpdf}
\end{align}

\noindent where 
$f_1$ and $f_2$ are the fractions for the composite PDFs of G (standard Gaussian), CB (Crystal-Ball~\cite{dscb}, with $\alpha$ and $n$ as parameters to model the high mass tail), and BfG (Bifurcated Gaussian with the widths $\sigma^{L}_3$ and $\sigma^{R}_3$ on the left and right side of the peak, respectively). All the three PDFs share a common peak position of $\left[\Delta m_{i}+\delta_{\Delta m_{i}}\right],\, (i\in \{+,s\})$, and a scale factor $\varepsilon$ for all the width parameters accounting for the resolution difference between data and MC simulation. 
To determine $\Delta m_+$ in the dedicated calibration channel of $D^{*+}\to D^+ (\to K^- \pi^+ \pi^+) \pi^0$, we first fit the signal PDF to correctly reconstructed signal MC events. In the fit, we fix $\Delta m_{+}$ to the generated value and $\varepsilon$ to one, while $\delta_{\Delta m_{+}}$ and other parameters are left free to vary. 
A subsequent fit to the data within a range of $\Delta M_+\in[0.135,\ 0.160]$~\gevcc, 
with $\Delta m_{+}$ and $\varepsilon$ free to vary and other signal PDF parameters fixed to the MC values, leads to $\Delta m_+=(140\,615.9 \pm 3.4)$~\kevcc\ with a signal yield of $72\,053\pm295$. Here the uncertainties are statistical only. 

The biases ${\rm \delta}_{\Delta m_{+}}$ and ${\rm \delta}_{\Delta m_{s}}$ obtained from the fits to signal MC samples may not be sufficient to cover all detector effects, especially those related to the electromagnetic calorimeter responses. The detector effects that are not modeled potentially will cause additional biases in measuring photon energies and directions. These biases will propagate into the $\pi^0$ momentum measurement, and eventually the measurements of $\Delta m_{s(+)}$. The biases in the $\pi^0$ momentum measurement are studied with real data in our calibration channel $D^{*+}\to D^+ (\to K^- \pi^+ \pi^+) \pi^0$ by measuring $\Delta m_+$ in the regions of the $\pi^0$ momentum $p(\pi^0)$ and the cosine of the polar angle of $\pi^0$ $\cos\theta (\pi^0)$, both in the laboratory frame. Here $\theta(\pi^0)$ is defined with respect to the $z$ axis, the symmetry axis of the multilayer drift chamber. Strong dependence of  $\Delta m_+$ as a function of $p(\pi^0)$ or $\cos\theta (\pi^0)$ is found.
As $p(\pi^0)$ and $\cos\theta (\pi^0)$ are highly independent of each other, a two-step correction procedure is adopted to correct the energy of $\pi^0$ first in five equally-sized intervals of $p(\pi^0)$ between 0 and 0.1~\gevc\ (Step1 correction), then in four intervals of $\cos\theta(\pi^0)$ between $-1$ and 1 (Step2 correction). At each correction step in each $p(\pi^0)$ or $\cos\theta(\pi^0)$ interval, $E(\pi^0)$ is changed by a certain amount $\delta_E$ for each $\pi^0$ candidate. The magnitude of the $\pi^0$ momentum, $p(\pi^0)$ is scaled accordingly, while its  direction remains unchanged. The variation of the $\Delta m_{+}$ central value after fitting the distribution of $\Delta M_{+}$ calculated with the updated $\pi^0$ four-momentum is found to depend linearly on $\delta_E$. 
After applying the correction to the signal MC sample of $D^{*+}\to D^{+}(\to K^-\pi^+\pi^+)\pi^0$ decays, the measured 
$\Delta m_+$ central value coincides with the input value $(140\,603 \pm 15)$~\kevcc~\cite{pdg} as expected,
and the dependence on either $p(\pi^0)$ or $\cos\theta (\pi^0)$ is found to be at minimal level. To validate our method, the same correction scheme is applied to the MC sample of  $D^{*+}\to D^{+}(K^+K^-\pi^+)\pi^0$ decays, and the measured $\Delta m_+$ in this decay channel is $(140\,589.8 \pm 12.8)$~\kevcc, with the data fit illustrated in the Supplementary material,
showing good agreement with the current world average within one standard deviation. 
As shown in Fig.~\ref{fig:fit_ds_double}, after applying the $E(\pi^0)$ correction scheme on the signal decay channel of \mbox{$\decaychainc$}, $\Delta m_s$ is determined to be $\Delta m_s=(144\,201.9\pm 44.2)$~\kevcc, where the uncertainty ($\sigma_{\rm stat}$) is statistical only.

The systematic uncertainties of $\Delta m_s$ are from various sources. The dependence of $\Delta m_s$ on $p(\pi^0)$, on $\cos\theta(\pi^0)$, on the azimuthal angle of $\pi^0$ meson, $\phi(\pi^0)$, and on $M(D_s^+)$ is studied in turn by collecting fit results for $\Delta m_s$ in subsets of data with roughly equal statistics for each parameter. Also, $\Delta m_s$ is measured in four non-overlapping regions of the $D_s^+\to K^+ K^- \pi^+$ Dalitz-plot. Furthermore, $\Delta m_s$  is measured in four disjoint subsets of data-taking periods. 
We adopt a method similar to the PDG scale factor~\cite{pdg,DSTpDpbabar2017} to determine our systematic dependence on a certain observable with variations larger than those expected from pure statistical fluctuations, as summarized in the Supplementary material.
A systematic uncertainty of 25.5~\kevcc\ is therefore assigned for the $M(D_s^+)$ dependence. 

In the nominal fit to data, nine signal shape parameters $\delta_{\Delta m_s}$, $\sigma_{1,2}$, $\sigma_3^{\rm L,R}$, $f_{1,2}$, $\alpha$ and $n$ are fixed to the values obtained from the fit to signal MC events. To account for the uncertainties and correlations of these parameters due to limited statistics of the signal MC events, 
we sample the signal shape parameters 1000 times based on the central values and the
covariance matrix from the fit to signal MC events. 
Each time, we refit the data with the parameters 
fixed to the sampled values. 
The distribution of the 1000 fitted central values of $\Delta m_s$ has a Gaussian width of $5.5~\kevcc$,  which is taken as the systematic
uncertainty for the signal shape parameters. We have additionally tested an alternative signal shape configuration of the sum of two PDFs, a CB and a BfG. The resulting change in the $\Delta m_s$ central value is negligible. 

To test whether the fit procedure introduces a bias in $\Delta m_s$, we perform closure checks with nine inclusive MC samples for the decay channel of \mbox{$D_s^{*+}\to D_s^{+}(\to K^+K^-\pi^+)\pi^0$} each with statistics matched to that in data. The fits yield a mean fit bias that is consistent with zero. Therefore, the systematic uncertainty related to the fit bias is negligible. 
\begin{figure}[t]
    \centering
    \includegraphics[width=0.97\linewidth]{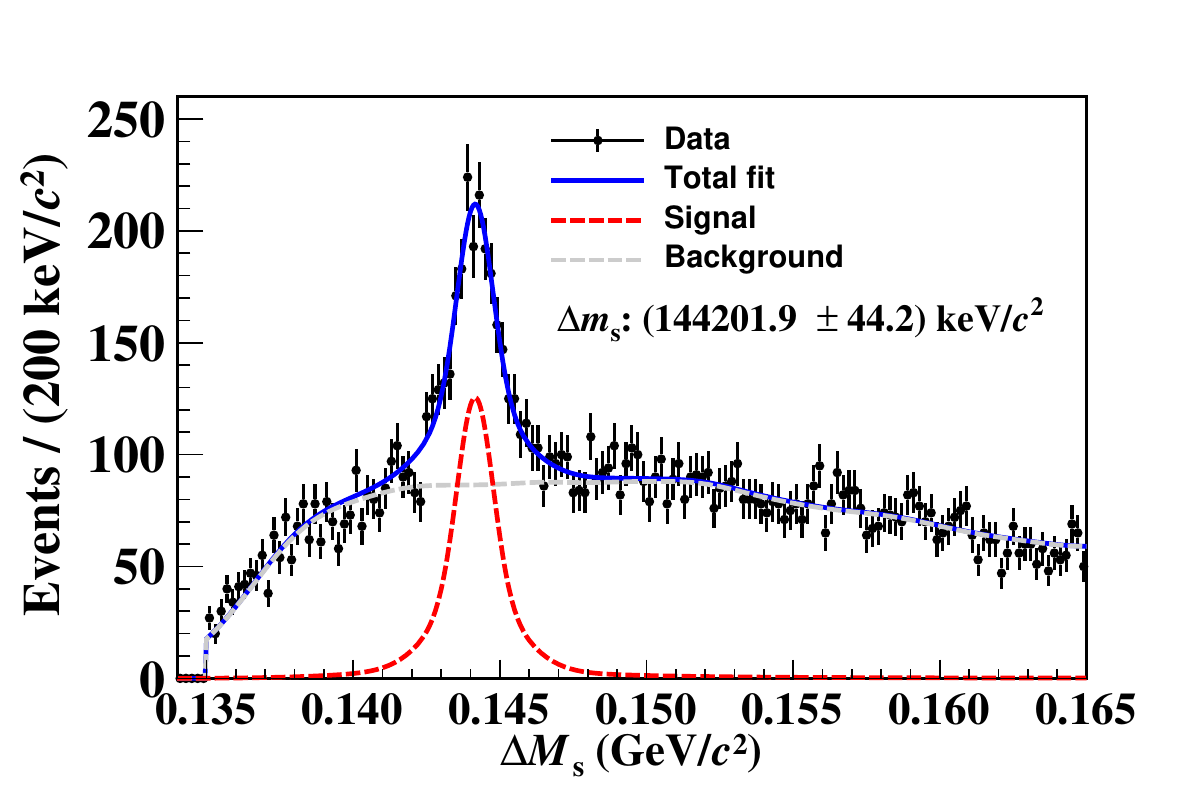}
    \caption{The fit to the $\Delta M_s$ distribution of the data sample of the decay chain $D_{s}^{*+} \to D_{s}^{+}(\to K^{+} K^{-} \pi^{+})\pi^{0}$. 
    }
    \label{fig:fit_ds_double}
\end{figure}

The systematic uncertainty from the background shape is studied by adopting an alternative background model to fit the data. This model is the sum of a kernel-estimation function obtained from MC simulation to  model peaking backgrounds that are predominately from $D^{*+}\to D^+\pi^0$ decays, and a threshold function~\cite{DSTpDpbabar2017,argus} to model random combinatorial backgrounds. 
The fit with the alternative background model leads to a change in the $\Delta m_s$ fit value that is smaller than its own statistical fluctuations, and hence no systematic uncertainty is assigned.

To account for the systematic uncertainty from the $\pi^0$ energy correction, we consider two different effects, one is due to limited statistics of the calibration sample of $D^{*+} \to D^{+}(\to K^{-} \pi^{+} \pi^{+})\pi^{0}$, the other is from the interval scheme on $p(\pi^0)$ and $\cos\theta (\pi^0)$. 
They are evaluated by adopting a series of alternative $\pi^0$ energy correction factors and interval schemes. We assign 9.1~\keVcc\ and 11.5~\kevcc\ as the systematic uncertainty related to the two effects, respectively.

To check the stability of the result with respect to the $M(\gamma\gamma)$, $M_{\rm rec}(D_s^{*+})$ requirements and the upper limit value of the $\Delta M_s$ fit range, we perform alternative fits with a series of corresponding parameter values. All the results are consistent with our nominal result, and no systematic uncertainty is assigned.

Moreover, our measurement of $\Delta m_s$ separately for $D_s^{*+}$ and $D_s^{*-}$ yields results compatible within one standard deviation. We also test that $\Delta m_s$ value is not sensitive to slight fluctuations of the magnetic field.

Combining all sources contributing to the systematic uncertainty in quadrature we find a total systematic uncertainty of 29.9~\kevcc\ for the $\Delta m_{s}$ measurement. A comprehensive table summarizing the systematic uncertainties is provided in the Supplementary material. The uncertainty due to limited study on $\Delta m_+$ relies on external input~\cite{pdg}, and is considered separately.

In summary, based on 3.19~$\ifb$ of $\ee$ annihilation data collected at $\sqrt{s}=4.178\gev$ with the BESIII detector, we have developed a novel data-driven $\pi^0$ momentum 
calibration technique 
and applied it to reduce the overall uncertainty on the
$ D_s^{*+} - D_s^+$ mass difference 
by a factor of seven (as shown in Fig.~\ref{fig:ex_com}).
Using $D^{*+} \to D^+ \pi^0$ decays as a 
control channel with the precisely known $D^{*+} - D^+$ mass difference as the anchor, 
we correct for detector effects in two-dimensional bins of $\pi^0$ momentum and 
polar angle, reducing systematic uncertainties to a level where the measurement 
remains statistically limited. 
This approach is directly applicable to other high
precision measurements involving soft neutral pion reconstruction. 
The measured mass difference,
$\Delta m_s = [144\,201.9 \pm 44.2\,(\mathrm{stat.}) \pm 29.9\,(\mathrm{syst.}) 
\pm 15.0\,(\mathrm{input})]$~keV/$c^2$,
differs from the chiral perturbation theory prediction of Ref.~\cite{Goity:2007fu}
by  $2.7\sigma$.
This result,
together with the $ \Delta m_0 $ and $ \Delta m_+ $
measurements made after that prediction,
should  motivate refined theoretical calculations. 
Similarly, the much more precise value of $ \Delta m_s $ 
presented here should lead to significantly more precise
lattice QCD calculations of the $ D^*_s$  width and decay constant
than were possible previously~\cite{Donald:2013sra}.
Finally, the 
derived $D_s^{*+}$ mass determination provides new input for 
better understanding 
the composition of $Z_{cs}$ exotic states near threshold, as measurements of their masses improve.

With the input uncertainty on $\Delta m_s$ 
canceled in the difference, we also report the measurement on the SU(3) violating parameter $\Delta m_D\equiv \Delta m_s-\Delta m_+=3.599\pm 0.055$~MeV/$c^2$. This is the most stringent test on SU(3) flavor symmetry for heavy-flavor hadrons. Considering the recently measured $\Delta m_B\equiv (m_{\bar B_s^{*0}}-m_{\bar B_s^{0}})-(m_{\bar B^{*0}}-m_{\bar B^0})=3.94\pm 0.14$~MeV/$c^2$ by the CMS collaboration~\cite{CMS:2025byz}, we have the ratio $\Delta m_D/\Delta m_B=0.91\pm 0.04$, that shows a clear deviation from the quark mass ratio $m_b/m_c$, indicating heavy-quark symmetry violation~\cite{Goity:2007fu,Karliner:2019lau,Meng:2022ozq}. More theoretical studies on heavy quark and chiral symmetries are needed to understand the strikingly small SU(3) violating effect of $\sim$2.5\% in the charm sector.\\

\begin{figure}[t]
    \centering
    \includegraphics[width=0.97\linewidth]{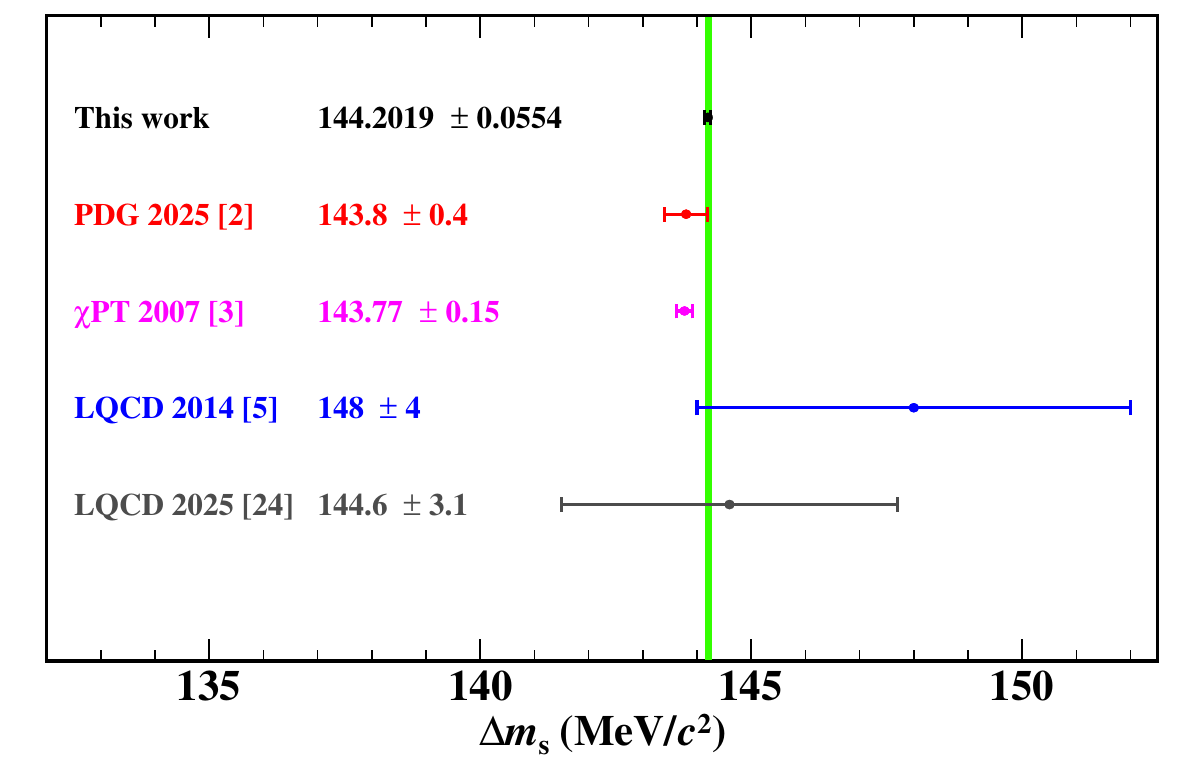}
    \caption{Comparison of $\Delta m_s$ obtained in this work with current world average~\cite{pdg}, the $\chi$PT-based prediction~\cite{Goity:2007fu}, and Lattice QCD calculations~\cite{Donald:2013sra,CLQCD:2024yyn}. The green band highlights our measurement with all uncertainties combined.}
    \label{fig:ex_com}
\end{figure}

\noindent
\textbf{Conflict of interest}\\

The authors declare that they have no conflict of interest.

\vspace{1cm}
\noindent
\textbf{Acknowledgments}\\

The BESIII Collaboration thanks the staff of BEPCII and the Institute of High Energy Physics (IHEP) computing center for their strong support. This work is supported in part by National Key R\&D Program of China (2023YFA1606000, 2023YFA1606704); National Natural Science Foundation of China (NSFC) (11635010, 11935015, 11935016, 11935018, 12025502, 12035009, 12035013, 12061131003, 12192260, 12192261, 12192262, 12192263, 12192264, 12192265, 12221005, 12225509, 12235017, 12361141819); the Chinese Academy of Sciences (CAS) Large-Scale Scientific Facility Program; the Strategic Priority Research Program of CAS (XDA0480600); CAS (YSBR-101); 100 Talents Program of CAS; The Institute of Nuclear and Particle Physics (INPAC) and Shanghai Key Laboratory for Particle Physics and Cosmology; ERC (758462); German Research Foundation DFG (FOR5327); Istituto Nazionale di Fisica Nucleare, Italy; Knut and Alice Wallenberg Foundation (2021.0174, 2021.0299); Ministry of Development of Turkey (DPT2006K-120470); National Research Foundation of Korea (NRF-2022R1A2C1092335); National Science and Technology fund of Mongolia; Polish National Science Centre (2024/53/B/ST2/00975); STFC (United Kingdom); Swedish Research Council (2019.04595); U. S. Department of Energy (DE-FG02-05ER41374).

\appendix
\section{Supplementary material}
\label{sec:appendix}
Supplementary data to this work can be found online at 

\href{http://doi.org/10.1016/j.scib.2026.04.059}{http://doi.org/10.1016/j.scib.2026.04.059}.

\begin{center}
	\textbf{BESIII Collaboration}
\end{center}
\begin{small}
M.~Ablikim$^{1}$\BESIIIorcid{0000-0002-3935-619X},
M.~N.~Achasov$^{4,b}$\BESIIIorcid{0000-0002-9400-8622},
P.~Adlarson$^{81}$\BESIIIorcid{0000-0001-6280-3851},
X.~C.~Ai$^{86}$\BESIIIorcid{0000-0003-3856-2415},
R.~Aliberti$^{38}$\BESIIIorcid{0000-0003-3500-4012},
A.~Amoroso$^{80A,80C}$\BESIIIorcid{0000-0002-3095-8610},
Q.~An$^{77,63,\dagger}$,
Y.~Bai$^{61}$\BESIIIorcid{0000-0001-6593-5665},
O.~Bakina$^{39}$\BESIIIorcid{0009-0005-0719-7461},
Y.~Ban$^{49,g}$\BESIIIorcid{0000-0002-1912-0374}
% ...
\\
\vspace{0.2cm}
(BESIII Collaboration)\\
\vspace{0.2cm} {\it
$^{1}$ Institute of High Energy Physics, Beijing 100049, People's Republic of China\\
% ...
$^{4}$ Budker Institute of Nuclear Physics SB RAS (BINP), Novosibirsk 630090, Russia\\
% ...
$^{38}$ Johannes Gutenberg University of Mainz, Johann-Joachim-Becher-Weg 45, D-55099 Mainz, Germany\\
$^{39}$ Joint Institute for Nuclear Research, 141980 Dubna, Moscow region, Russia\\
% ...
$^{49}$ Peking University, Beijing 100871, People's Republic of China\\
% ...
$^{61}$ Southeast University, Nanjing 211100, People's Republic of China\\
% ...
$^{63}$ State Key Laboratory of Particle Detection and Electronics, Beijing 100049, Hefei 230026, People's Republic of China\\
% ...
$^{77}$ University of Science and Technology of China, Hefei 230026, People's Republic of China\\
% ...
$^{80}$ University of Turin and INFN, (A)University of Turin, I-10125, Turin, Italy; (B)University of Eastern Piedmont, I-15121, Alessandria, Italy; (C)INFN, I-10125, Turin, Italy\\
$^{81}$ Uppsala University, Box 516, SE-75120 Uppsala, Sweden\\
% ...
$^{86}$ Zhengzhou University, Zhengzhou 450001, People's Republic of China\\
\vspace{0.2cm}
$^{\dagger}$ Deceased\\
% ...
$^{b}$ Also at the Novosibirsk State University, Novosibirsk, 630090, Russia\\
% ...
$^{g}$ Also at State Key Laboratory of Nuclear Physics and Technology, Peking University, Beijing 100871, People's Republic of China\\
% ...
}
\end{small}

\end{document}